# The 'crisis of noosphere' as a limiting factor to achieve the point of technological singularity

Rafael Lahoz-Beltra

Department of Applied Mathematics (Biomathematics). Faculty of Biological Sciences. Complutense University of Madrid. 28040 Madrid, Spain. lahozraf@ucm.es

## 1. Introduction

One of the most significant developments in the history of human being is the invention of a way of keeping records of human knowledge, thoughts and ideas. The storage of knowledge is a sign of civilization, which has its origins in ancient visual languages e.g. in the cuneiform scripts and hieroglyphs until the achievement of phonetic languages with the invention of Gutenberg press. In 1926, the work of several thinkers such as Edouard Le Roy, Vladimir Vernadsky and Teilhard de Chardin led to the concept of noosphere, thus the idea that human cognition and knowledge transforms the biosphere coming to be something like the planet's thinking layer. At present, is commonly accepted by some thinkers that the Internet is the medium that brings life to noosphere. Hereinafter, this essay will assume that the words Internet and noosphere refer to the same concept, analogy which will be justified later.

In 2005 Ray Kurzweil published the book *The Singularity Is Near: When Humans Transcend Biology* predicting an exponential increase of computers and also an exponential progress in different disciplines such as genetics, nanotechnology, robotics and artificial intelligence. The exponential evolution of these technologies is what is called Kurzweil's "Law of Accelerating Returns". The result of this rapid progress will lead to human beings what is known as technological singularity. According to Vinge (2013) and Kurzweil's technological singularity hypothesis, noosphere would be in the future the natural environment in which "human-machine superintelligence" emerges after to reach the point of technological singularity.

According to James Lovelock's Gaia hypothesis (Figure 1) the living and non-living parts of the planet form a self-regulated complex system maintaining life on Earth (Lovelock, 1979). Such whole system is known with the name of biosphere. Somehow the living and non-living beings evolve together (Weber, 2001; Nuño et al., 2010), having organisms an influence on their environment (Watson and Lovelock, 1983) and the environment in turn affects the organisms by means of Darwinian natural selection. For instance, photosynthetic organisms regulate global climate, marine microorganisms may be keeping the oceans at a constant salinity and nitrogen-phosphorus concentrations, etc. In agreement with Vernadsky (2004) a prerequisite for the coming of the noosphere is the existence of the technosphere. In some way the biosphere is a stable, adaptive and evolving life system with sufficient free energy to power the launching of a technosphere (Lahoz-Beltra, 2008). Therefore, technosphere emerges as a physical environment on Earth being a new layer inhabited by machines, cities and industry with an influence into the biosphere (Figure 1).

However, and according to the data available today, how realistic is the technological singularity hypothesis? In this essay we present a criticism of Kurzweil's "Law of Accelerating Returns" focusing on the fact that the exponential growth assumes unlimited resources and energy. Our criticism of Kurzweil's ideas is inspired by computer video games simulating the course of a civilization or a city, and the predictions obtained with simple simulation experiments using differential equation models of population dynamics.

In this essay we show that if we consider the energy that sustains the noosphere, i.e. Internet, and simulating its growth by means of an exponential numerical model it is impossible that our civilization reaches the point of technological singularity in the near future. The model is based on some fundamental assumptions and simple simulation experiments, obtaining as a plausible scenario what we have called as 'crisis of noosphere'. Assuming Internet stores today at least 1000 Exabytes (1 Exabyte = $10^{18}$ bytes) and human knowledge doubling occurs every 12 months, will come a point in the next 50 years (by the year 2063) or maybe before when Internet will consume the total electricity produced worldwide. With current technology this energy would be equivalent to the energy produced by 1,500 nuclear power plants. Once this happens there will be a collapse of the noosphera and possibly part of the biosphere. Therefore, we believe that with the current technology we are really far from reaching the point of technological singularity.

However, we believe that if a 'paradigm shift' occurs first then the singularity point could be reached later. Thus, the point of singularity could be achieved with a paradigm shift, namely the design of a noosphere which hardware be adaptable to available energy and designing a more efficient computer machines that the current ones. A sistemic model of a noosphere ranging in size depending on the available energy is simulated according to the Lotka–Volterra equations, assuming that the Internet is a *predator* specie that feeds voraciously on a *prey*, the electric power. A hardware architecture with this dynamic behavior would allow an Internet based on 'computer machines' more effective in terms of power consumption than current ones. And in this respect the Volterra-Lotka model could give us some clues about how the Internet should be designed.

Moreover, we propose the use of non-conventional technologies for the design of a new class of computer-oriented to the implementation of the noosphere. In this essay we speculate about what we have called 'Nooscomputer' or *N-computer*, a hypothetical machine, resembling a Turing machine, but capable of storing and processing human knowledge through quantum computation, DNA and Egan's *Permutation City* algorithms. The use of N-computers in data centers would allow a new class of Internet which would consume far less power allowing our civilization to reach the point of technological singularity.

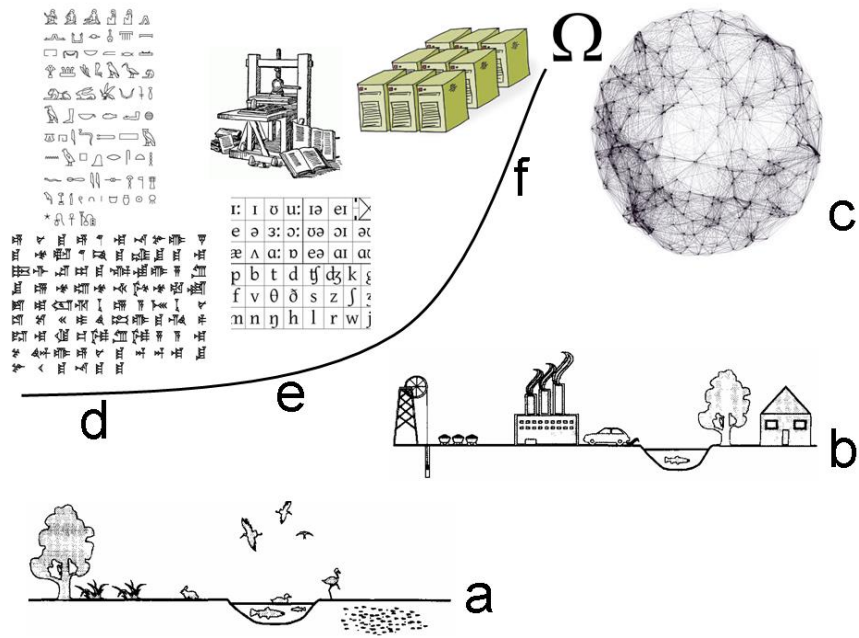

**Figure 1**. Vernadsky's hypothesis of Gaia states that noosphere (**c**) is the third layer of development of the Earth, after biosphere (**a**) and technosphere (**b**). For Vernadsky noosphere is "the last of many stages in the evolution of the biosphere in geological history". Teilhard de Chardin states that noosphere is growing towards the Omega point ( $\Omega$ ). Initially the ancient noosphere was very primitive and knowledge was stored in stone or papyrus (**d**, cuneiform scripts and hieroglyphs), later in paper (**e**, Gutenberg press) and currently as a global network of computers (**f**, Internet). Therefore, with the passage of time has changed the information storage media. Now, we must ask the question, is the noosphere energetically sustainable?

## 2. Is the Internet the nervous system of the Noosphere?

Noosphere is a term that was introduced by Édouard Le Roy (1870–1954), Vladimir Vernadsky (1863-1945) and Teilhard de Chardin (1881-1955) referring to the sphere of human thought (Levit, 2000). Edouard Le Roy was the first who used the notion "noosphere" in a scholarly publication entitled *L'exigence idéaliste et le fait l'évolution* published in 1927. Le Roy (1927) wrote:

"Beginning from a man, evolution carried out with new, purely psychic meanings: through the industry, society, language, intellect etc, and thus biosphere transforms into noosphere".

However, the explanation of how it arises varies from one thinker to another. It was coined by the French theologian and scientist paleontologist Pierre Teilhard de Chardin in 1925 and disseminated in posthumous publications in the 1950s and 1960s. According to Teilhard (Teilhard de Chardin, 1959, 1964) the noosphere emerges as a result of the interaction among humand minds. However, for Vernadsky (2004) and although it is not a material layer the noosphere emerges once of human beings sufficiently progress, for example reaching the ability to create new material resources. In the point of view adopted by Teilhard the noosphere will grow up to the point called Omega, the end of history (Figure 1). For Teilhard the Omega point is the maximum level of complexity and consciousness towards which he believed the universe is evolving. The evolution of humanity toward some ideal situation has received several names, whether Omega point or technological singularity. But whereas the first term has a spiritual meaning the second one has a technological taste. The concept of technological singularity was introduced in the 50s by John von Neumann (Ulam, 1958) who thought that humanity is approaching some a point where once reached would change the course of humanity. At present the singularity advocates predict an explosion of intelligence in which powerful supercomputers and other machines will be well over human skills and intelligence. Among other factors, this explosion of intelligence will result from a dramatic breakthrough in artificial intelligence. So while the concept of Omega point is characteristic of theistic evolutionists, e.g. Francis Collins (Collins, 2006), the concept of technological singularity usually is defended by scientists for whom science promises a limitless evolution of humankind. No matter how this critical point is designated, in this essay we will denote it by $\Omega$. At presente there are several predictions of possible dates in which our civilization will reach this point. All dates have been proposed have one feature in common, either 2045 predicted by Kurzweil or 2030 predicted by Vinge: in just a few years we will have reached the $\Omega$ point.

At present, there are several opinions that support the role of Internet as the nervous system of the noosphere. For example, Hagerty (2000) thinks that Internet is playing the role that Teilhard termed "the mechanical apparatus" of the noosphere. Shenk (1997) believes that even when the World Wide Web is a repository for the knowledge of humankind, it is only the beginning of the development of the global mind, and therefore the noosphere. According to Heim (1993), Teilhard envisioned the convergence of humans in a single massive "noosphere" or "mind sphere", maybe the "virtual communities" mentioned by McLuhan's global village or Teilhard's Omega point. For this thinker and philosopher concerned with virtual reality we have enriched the process of creating further realities through virtualization. If we take another step in its reasoning assuming that susch virtual realities are "windows to the noosphere" we arrive at the conclusion that the hardware of such virtualization is the Internet.

However, how the noosphere originates from the technosphere? In this essay, we propose the following model. According to Figure 2 technosphere would be an open system since it meets the following characteristics: (i) Consists of a set of parts which interact, (ii) it is oriented to a purpose, (iii) consumes materials, processes it to produce a product or service, (iv) consumes energy, (v) interacts, reacts and affects the environment, (vi) it grows, changes and adapts, that is evolves and finally (vii) it competes with other systems, for example with the biosphere. Accordingly as such open system technosphera would have an *input* (information, money, enery, resources, etc.) and *output* (information, money, goods, services, etc.). The in-put processing would results in several outputs, including information. The technosphere also produces other undesirable outputs, such pollutants and wastes. Then a part of or all the information obtained can be processed again becoming knowledge. Transforming information into knowledge means that technosphere was able to discover patterns, relationships and trends resulting in formalized and objective contents. Therefore, while the information may be stored in a database, the knowledge requires more sophisticated media, for example in the knowledge base of an expert system (Lahoz-Beltra, 2004). Another possibility is that the information is forwarded to the input to be processed again. Finally, as the technosphere produces information and knowledge, these are embodied in a new entity: the noosphere.

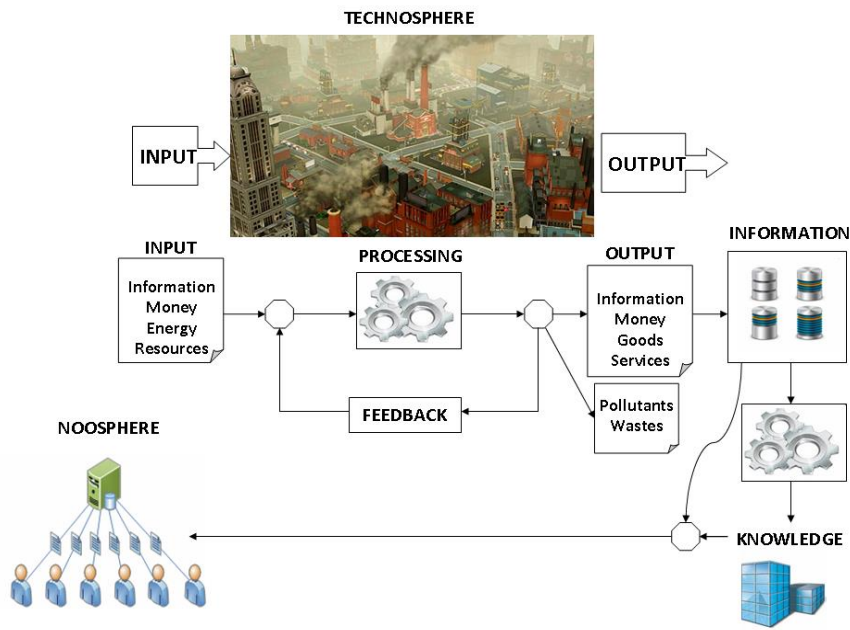

**Figure 2**. A possible explanation of the origins of the noosphere from the technosphere (for explanation see text).

## 2.1. The origins of the crisis: technosphere complexity and energy consumption

In accordance with previously described model the noosphere is arising from development of technosphere. This means that the noosphere inherits all the strengths and weaknesses of its predecessor, the tecnosphere. Consequently, what are the causes that lead to a crisis of the noosphere? Consider the following possible explanation.

Recently Arbesman[1] used *SimCity* -an open-ended city-building computer video game- to measure a city's Kolmogorov complexity. Thus, the complexity of a city could be measured as the shortest algorithm required to reproduce it. Since the technosphere includes the cities of our planet Earth, this method could be used to estimate a minimum value of complexity of the technosphere from the complexity value obtained in the cities. In a theoretical realm, using a small dataset of population sizes and file sizes of some cities constructed in SimCity 3000, Arbesman found how complexity linearly scales with population size. Furthermore, it is interesting to note how this result coincides with similar others obtained with real cities. Bettencourt et al. (2007) demonstrated as most aspects of a city, such as electrical use, employment, or population growth, increase linearly based on the size of that city.

Within this framework human societies are distinguished by their dominant pattern of energy harvesting, a behavior which has been called the *energy paradigm* (White, 1959; Karakatsanis, 2010). For example, Garrett (2011) modeled the civilization as 'heat engine' which requires energy consumption. In accordance with this paradigm collapsing civilizatios are complex systems that continued to grow beyond the limits of their energy budget. This would be true unless such civilization makes the effort to find an efficient mechanism of technological transition. Therefore, there is a limit to the unlimited growth of the technosphere unless we are able to find a new technology with which to build a new technosphere energetically more efficient and therefore sustainable. That is, the 'crisis of noosphere' would be a consequence of the high energy consumption of the technosphere, and therefore the noosphere would inherit from technosphere this major flaw. Today, Internet - the nervous system of the noosphere - has become a true electric power predator.

---

[1] See S. Arbesman. 2012. *The mathematical puzzle that is the complexity of the city*. http://www.theatlanticcities.com/politics/2012/06/mathematical-puzzle-complexity-city/2261/

[2] See C. Keenan. 2011. *How much energy does the Internet consume?* http://planetsave.com/2011/10/27/how-much-energy-does-the-internet-consume/

## 3. A numerical model of Noosphere

Numerical models are mathematical models that use some sort of numerical time-stepping procedure for predicting the system behavior over time. One application of the numerical models is the study of complex societies as a predictable phenomenon, making predictions according to a mathematical model. Whether using differential equations (Good and Reuveny, 2009; Griffin, 2011) or probability theory (Gavrilets et al., 2010), it is possible to explain from collapse societies such as the Maya to the effects of warfare or some social policies. In fact, some popular computer games, for example *Civilization V*, behave according to some differential equations such as the Malthusian model and simple polynomials of different degrees (ref.: see in Internet the blogs about this game).

In this section and using an exponential or Malthusian growth model [1], we build a simple model (see Appendix, *Scenario 1*) that illustrates what we have called as 'crisis of noosphere':

$$\dot{y_1} = k_1 y_1 \quad [1]$$

where $y_1$ is the amount of information stored in Internet, the noosphere, and $k_1$ is the Malthusian parameter, thus the information growth rate.

Based on the reasoning of the previous section, consider the following facts about Internet, the hardware that gives life to the noosphere:

1. Digital information is housed in data centers around the world, doubling in size every 2 years. There are over 500,000 data centers in the world[2].

2. Internet and other forms of information technology account for 2% of all electrical energy used globally. Most data centers that house computer servers rely on non-renewable energy resources, i.e. nuclear and coal-powered energy[2].

3. Estimates by US industry experts: Internet uses 30 billion watts or 30 nuclear power stations[3].

4. Moore's law can be applied to the amount of information people add to the Internet each day. The US alone is home to 898 exabytes (1 EB = 1,000 Petabytes = 1,000,000 Terabytes = 1 billion gigabytes) - nearly a third of the global total information (Western Europe has 19% and China has 13%)[4].

Assuming that digital information housed in data centers is doubling its size every 2 years, $k_1$=2, and setting an initial amount of information $y_0$ equal to 1000 EB, likely an underestimation of the global actual value, we obtained the results shown in Figure 3:

---

[3] See M. Tyson. 2012. *The Internet uses 30 nuclear power station's energy output*. http://hexus.net/tech/news/industry/45689-the-internet-uses-30-nuclear-power-stations-energy-output/

[4] See S. Forman. 2013. *The US is home to one third of the world's data – here's who's storing it*. http://qz.com/104868/the-us-is-home-to-one-third-of-the-worlds-data-heres-whos-storing-it/

$$y = y_0 \exp(k_1 t) \quad [2]$$

Since 30 nuclear plants today represent 2% of the electrical energy consumed globally, we deduce that by the year 2062 noosphere will require about 1,500 nuclear power plants, this is 100% of the overall electric energy produced on Earth. According to Figure 3 from the year 2030 the amount of information stored in the noosphere will be around of $10^{17}$. This amount of information will have almost doubled the age of the universe ($10^9$ according to NASA'S WMAP Project) and account for 21% of the number of atoms in the universe ($10^{78}$).

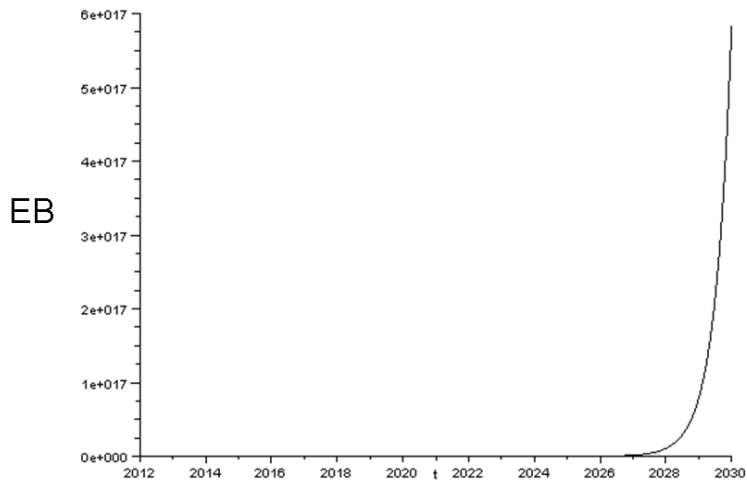

**Figure 3**. Exponential growth of the size of the noosphere (EB) under the assumptions of the "technological singularity" hypothesis.

Is this a technologically sustainable scenario? definitely not. According to Theodore Modis[5] in his excellent essay "The Sigularity Myth", there are several reasons why the "singularity" is not so near. We are not saying that it is unreachable in the future, but it is impossible to reach with current technology. Among all these reasons the most important is that (*i*) all natural growth follows the S-curve or logistic function which can be approximated by an exponential or Malthusian model in its early stages. For instance, Modis show how world population has grown significantly during the 20th century depicting an exponential model during early decades which becomes a logistic after World War II. A similar behavior can be seen with the cumulative oil production in the US and Moore's law, concluding that all exponential natural phenomena will eventually turn into logistics (Modis, 2003). Therefore, if the exponential is replaced by an S-curve, what effect will have this change in the predictions made by Kurzweil?

One of the Kurzweil's predictions is that the singularity will be reached by 2045. However, in agreement with our model (*ii*) to this date and if our civilization continues with a similar technology to current, the noosphere will store an amount of information equal to $4,295 \times 10^{12}$. On that date the noosphera will consume a 66% of the overall electric power produced on Earth or equivalently the electrical energy produced by 990 nuclear power plants.

In agreemen with Modis (*iii*) the date on which the singularity is reached is strongly dependent on the evolution of the performance of computational power. Moreover, we think that the von Neumann architecture of current computers is not the most suitable to manage information and knowledge. For this reason we believe that our civilization have to face in the near future a *technological transition* which will result a new computer architecture. In this sense another of the predictions made by Kurzweil states that assuming tha the exponential trend will continue until 2045 the computer power will reach $6 \times 10^{23}$ Flops (floating-point operation per second). However, according to Modis and assuming a S-curve after 2045 computers will reach a maximum value of $10^{25}$, a value which clearly contradicts the Kurzweil's prediction of $10^{50}$ and above.

---

[5] See T. Modis. 2006. *The singularity myth*. http://www.growth-dynamics.com/articles/Kurzweil.htm#_ftn1

The scientific and technological criticisms mentioned above (*i, ii* and *iii*) seem sufficient to justify the need for a technological transition, which we believe should take place as a prior condition to the singularity point. Therefore, is our present civilization ready to face this technological change? At present we know that (*iv*) technological breakthroughs emerge a similar manner to the evolution of species, what is known as the punctuated equilibrium principle. This principle introduced by the naturalist Stephen Jay Gould states that speciation occurs in spurts of major changes that punctuate long periods of little change. According to theoretical predictions (Modis, 2003) in the case of technological breakthroughs the future milestones will appear progressively less frequently. In fact, there are thinkers like Huebner (2005) who argues that the rate of technological innovation is at present decreasing. For instance, the number of patents has been declining since the period 1850-1900.

In sumary, if we want our civilization at some time reaches the singularity point we will have to change the technology that currently sustains the noosphere.

## 3.1. A systemic model of the noosphere

In this section we will propose an alternative numerical model for the noosphere. Although the model is actually a metaphor, it may help us to find the conditions under which the Internet would be energetically sustainable. Inspired by the ecology we assume in the model that the Internet is a *predator* specie (for example foxes) that feeds voraciously on electric power. Hence, we assume that electrical energy represents the other specie, specifically the *prey* (for example rabitts). The Lotka–Volterra equations [3] (Lotka, 1925; Volterra, 1926) arise when the predator $y_2$, thus Internet, is related with the prey, the electrical energy $y_1$, occurring the coexistence of both species (Figure 4):

$$y_1^{'} = y_1 \left( k_1 - k_3 y_2 \right)$$

$$\text{[3]}$$

$$y_2^{'} = y_2 \left( -k_2 + k_3 y_1 \right)$$

where $k_1$, $k_2$ and $k_3$ are parameters describing the electric energy production, the loss rate of noosphere (in size) and the interaction (predation rate) of the two species, respectively. Equations [3] allow us to have a systemic model of the noosphere since this shares some characteristics with other phenomena in which the equations have been applied, e.g. predator–prey interactions, in the theory of auto-catalytic chemical reactions as well as in economic theory and modeling historical civilizations (Good and Reuveny, 2009).

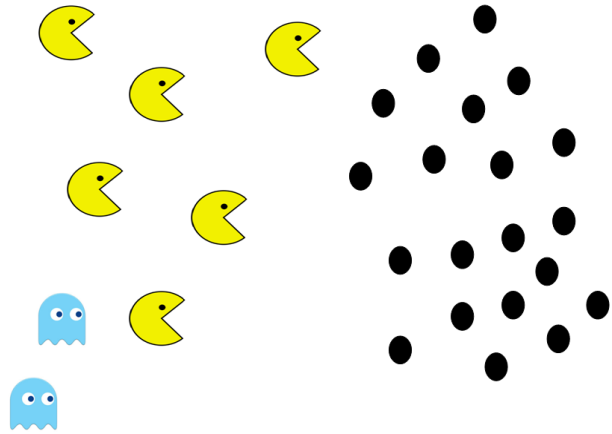

**Figure 4**. 'Pacman game' without ghosts (they eat the Pacs) is a good metaphor for a systemic model of the noosphere (the dots represent the energy/prey, the Pac represents Internet/predator). In this case the number of Pacs and dots available could coexist in equilibrium according to a Volterra-Lotka model.

However, since this model is a metaphor, indeed a thinkertoy, it is assumed that the size of the noosphere (e.g. EB, number of data centers, servers etc.) varies. Of course, the variable size of the noosphere is an Internet feature that should be included in the future to make it energetically sustainable. That is, its size varies according to the increasing or decreasing of the amount of electrical power available on Earth. Also we assume that the amount of electric energy is measured, e.g. as number of nuclear power plants. The number of power plants available varies according the size of the noosphere. Using the model [3], parameter values and initial conditions (see Appendix, *Scenario 2*) we illustrate a plausible systemic model of the noosphere (Figure 5).

Following, we propose an alternative model in which the electrical energy equation is modified according to the energy paradigm. According to Karakatsanis (2010) the dynamics of an energy paradigm could be expressed by the following equation:

$$y_1(t) = \varepsilon_t \, y_1(t-1) - \frac{(1-\alpha).\varepsilon_t}{A_t} y_1^2(t-1) \quad [4]$$

In equation [4] the ratios between the parameters $\varepsilon$, $\alpha$ and $A$ define the model dynamics. Inspired by this expression we included a $k_4 y_1^2$ term in the first equation of the Lotka–Volterra model [3] simulating a civilization, for example our civilization, evolving its harvesting energy policy under the pressure of resource depletion:

$$y_1^{'} = y_1 \left( k_1 - k_3 y_2 \right) - k_4 y_1^2$$
$$y_2^{'} = y_1 \left( -k_2 + k_3 y_1 \right)$$
$$[5]$$

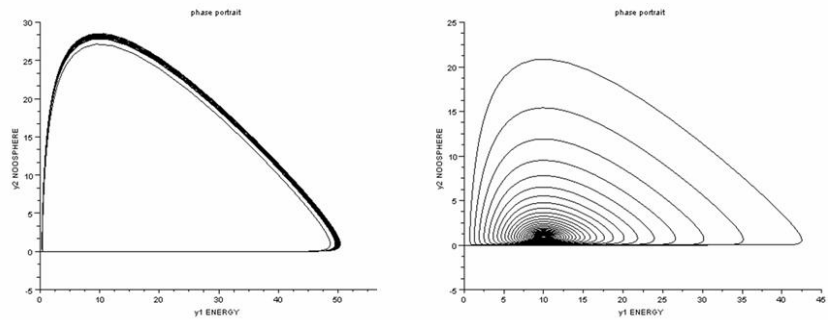

**Figure 5**. Cycles solution for coexistence between noosphere and electrical energy. (Left) *Scenario 2*. (Right) *Scenario 3*.

Using these new expressions we can simulate a new scenario for the noosphere (see Appendix, *Scenario 3*) under the influence of so-called energy paradigm. In the model [5] $k_4$ is the resource depletion pressure, i.e. electric energy.

## 5. But is there any possibility of achieving $\Omega$? Thinking about *N-computers*

As it mentioned in the introduction in this essay we speculate about *N-computers*, an abbreviation for 'Nooscomputer', thus a hypothetical machine resembling a Turing machine but capable of storing and processing human knowledge. At present such machine could be imagined as a result of very diverse technologies, namely through quantum computation, DNA and Egan's *Permutation City* algorithms. One goal in this essay is to give some highly speculative solution to criticisms that we discussed earlier.

The use of N-computers in data centers would have two advantages. On one side, future generations will have a new class of Internet with (*i*) electric power *consumption* well below the current one, on the other hand an Internet designed according to a (*ii*) *scalable architecture*, i.e. the number of servers would increase or decrease depending on knowledge storage needs and electric energy available.

Figure 6 shows a sketch of the elements and logical organization of N-computers. At right, and inside of a box, the sketch shows the main memory, control unit (CU), registers (AX, BX) and arithmetic logic unit (ALU). Although in principle this architecture is similar to the current computers based on von Neumann architecture, our proposal varies significantly by the 'hardware' from which elements are made. For example, whereas the main memory is DNA, the ALU operates based on cellular automata. Thus, the microprocessor is a cellular automata engine combined with DNA-based computing. On the left, the figure shows how the N-computer is provided with a von Neumann self-replicating automaton[6]. This automaton would allow the self-replication of N-computer like a predator specie growing Internet, and therefore the noosphere, according to Volterra-Lotka model. Thus, while a significant portion of the DNA memory is dedicated to storing knowledge, a small portion is dedicated to encode the information for self-replicating a copy of the N-computer. Of course, an N-computer also can be killed, i.e. shutdown, if required by the Volterra-Lotka dynamics.

---

[6] It is a self-replicating machine designed by John von Neumann. Using a construction arm or a writing head the machine can construct or print out a new copy of itself. The sequence of operations to be performed by the machine are encoded into a 'tape', i.e. DNA memory. A very interesting idea relating this self-reproducing automaton (universal constructor) with the ALU automata depicted in Figure 7 (top), is the existence of Garden of Eden configurations. However, we will not explain here this notion to be outside the scope of this essay.

In 2012 Church et al., (2012) stored a few petabytes ($10^{15}$) in a single gram of DNA. They encode one bit per base: adenine (A) or cytosine (C) for zero, guanine (G) or thymine (T) for one, synthesizing strands of DNA that stored 96 bits. In order to read the data stored in DNA the sequence of bases (A, T, G, C) is translated to a binary string. Each strand of DNA has a 19-bit address sequence playing the role of a memory address. One of the most interesting features of the 'DNA memory' is its great stability. Nearly a year after this finding Goldman et al. (2013) were able of encoding all 154 of Shakespeare's sonnets in DNA, 26-second audio clip from Martin Luther King's famous "I have a dream" speech and a copy of James Watson and Francis Crick's classic paper on the structure of DNA.

Based on previous experiments we propose that the N-computer main memory would be developed with DNA strands. Registers (labeled in the sketch as AX and BX, obviously N-computers can have more than two) could be implemented too with small DNA memories while the CU is a mixture of DNA and enzymes[7]. The CU performs very different functions, either converting a DNA sequence to binary, the inverse operation, or such as search engine (maybe a version of Google at molecular level or 'Moolgle') to identify certain sequences in the DNA strand. CU also controls which DNA sequences are read, written or deleted, ruling an important mission: it includes the molecular machinery necessary for DNA replication. The latter task is required when von Neumann self-replicating automaton comes into operation being obtained a child N-computer.

Consider the following example. If a DNA sequence is TATAGCCG storing some knowledge about the Roman Empire, establishing that the subunits of DNA T and G are equal to 1 and adenine A and C are 0, then CU transforms this sequence to 10101001. This sequence in binary is temporarily stored in the AX register and from here defines the initial state of the cellular automaton in the ALU. Applying this procedure, the N-computer can perform the processing of knowledge or conduct operations with the knowledge storaged in the DNA, e.g. change data, delete data, relate two DNA sequences storing related knowledge, etc.

---

[7] Enzymes are proteins that catalyzes a chemical reaction transforming a molecule or substrate to a new molecule or product (Recio Rincon et al., 2013).

In this new computer design how could be implemented the ALU?

In 1994 Greg Egan wrote *Permutation City* a hard-core science fiction novel that explores a model of consciousness and reality. Despite being a science fiction novel the computational paradigm that underlies is extremely interesting and suggestive. The author assumes that consciousness is Turing computable and consequently it could be simulated by a computer program. Thus, consciousness could be computed with a very simple machine which is restricted to a few simple operations, named Turing machine. In addition it is also assumed that it is possible to "copy" the consciousness of a human brain, "living" these copies or brain emulations as objects in a virtual reality (VR) environment. From these assumptions a VR city is created, Permutation City. In this VR environment copies are the only objects simulated in full detail, while the remainder of the objects are simulated with varying the rendering grain, using lossy compression and hashing algorithms[8]. At one point the Egan's novel explains that the city is a fragment of a Garden of Eden[9] configuration of an expanding massively celullar automata (Wolfram, 2002; Lahoz-Beltra, 2004) universe known in this fiction as TVC (Turing, von Neumman, Chiang). Since Garden of Eden configurations can only be obtained if the simulation has been designed for this purpose by an intelligent being, then this configuration is used as clue to show that a copy is 'living' in a simulated world. However, and this is the scientifically interesting aspect of the novel, this cellular automata universe has properties that make it similar to *spacefiller* configuration in Conway Game of Life[10]. Consequently, we will design the ALU of N-computers taking inspiration from this novel.

---

[8] Lossy compression algorithms are multimedia data encoding methods (audio, video, images), e.g. JPEG, MP3. Hash functions are procedures for transforming data of variable length to data of a fixed length, e.g. using http://www.fileformat.info/tool/hash.htm we transformed the text *Singularity hypothesis* to MD5 *a51129b92a02a1f932e63ce0ea586381*.

[9] It is a cellular automata pattern or configuration that has no parents or a predecessor configuration. Fort instance, in the Game of Life the pattern depicted in Figure 7 (top).

In an N-computer we replace the 'consciousnesses by knowledge'. An important feature of this change is that whereas consciousness is not computable (Hameroff, 1998), and therefore can not be simulated by a Turing machine as in the Egan's novel, knowledge is computable and therefore treatable with a Turing machine. An example of the latter are expert systems (Lahoz-Beltra, 2004).

Since the binary sequences representing some knowledge, e.g. 10101001, are the input data or initial cellular automata configuration of the ALU it seems appropriate that ALU performs operations based on something similar to some configurations found in Conway Game of Life. This procedure is what we call in this essay as *Egan's algorithm*. Finally, the output will be the final state of the automata (at equilibrium, or at a given iteration, etc.) which can be transformed from a binary code (0s and 1s) to a DNA sequence of A, T, G and C.

One of the hard problems to be solved on the N-computers is their construction. Although it may result amazing some authors have been able to combine DNA, quantum computing and cellular automata in the same recipe (Goertzel, 2011): the *femtocomputing* paradigm. De Garis (2011) shows theoretically how the properties of quarks and gluons can be used (in principle) to perform computation at the femtometer (i.e. $10^{-15}$ meter) scale. Therefore, an N-computer could be made using non-standard hardware, thus with unconventional computational hardware.

---

[10] It is a cellular automata introduced by John Horton Conway in 1970. Once initial configuration is created (the only input), the player only observes its subsequent evolution. Each cell in the grid can be live (state 1) or dead (state 0). A live cell with two or three live neighbors stays alive and has a neighborhood consisting of the eight cells. The state of the cells is updated according to the following rules: (i) A dead cell (0) with 3 live neighbors becomes a live cell (1); (ii) a live cell (1) with 2 or 3 live neighbors stays alive (1); (iii) in all other cases, a cell dies or remains dead (0). A spacefiller is any pattern that grows at a quadratic rate by filling space with a periodic configuration in both space and time. For instance, in the Game of Life the pattern depicted in Figure 7 (bottom).

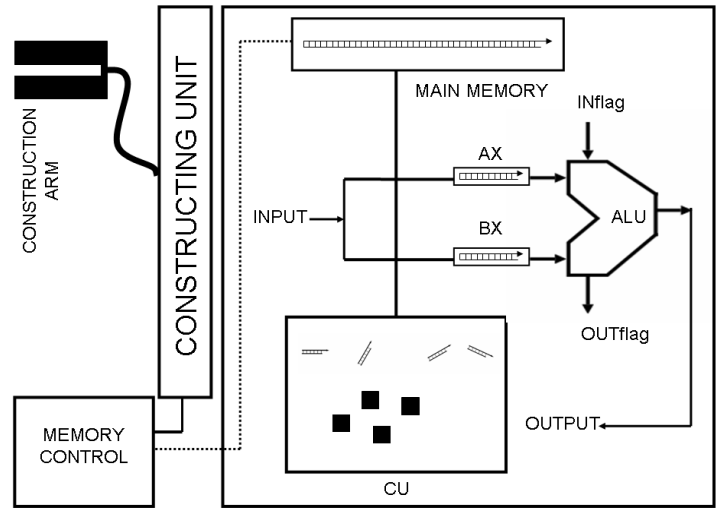

**Figure 6**. N-computer architecture (for explanation see text). In the 'Pacman game' – Figure 4- each Pac would have a skeleton as shown in this sketch.

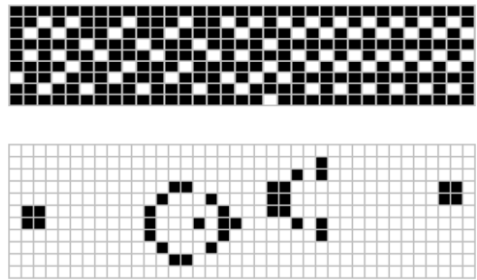

**Figure 7**. Garden of Eden (top) and spacefiller (bottom) configurations.

# 6. Conclusions

In 1962 Arthur C. Clarke wrote the novel *Profiles of the Future: An Inquiry Into the Limits of the Possible*, writing the following thought:

> "Any sufficiently advanced technology is indistinguishable from magic."

Thirty-two years later Greg Egan wrote *Permutation City*, saying at some passage in the novel:

> "Computers aren't made of *matter*".

At present human beings from most developed countries possess smart phones, laptops, tablets and other inventions with which they communicate with other humans or receive information on news, weather or predictions about the stock market. All these gadgets use to communicate with each other and they feed on the invisible layer that stores the information and knowledge, the noosphere. The speed of communication and the fact that the complex mechanisms that enable this technology are not transparent to the user, have led to the false impression that these inventions are a kind of magic. However, and for the same reason that genetic information requires a material substrate, DNA, the noosphere and all the gadgets that inhabit it, also require a substrate, Internet. With current technology the Internet has become a true electric power predator and it is for this reason that we see difficult to achieve in a few decades the $\Omega$ point. For this reason, and if we ignore the energy paradigm is easy to fall into the trap of assuming exponential models of resources, energy and space available, which are not unlimited. A simple model like the one we have called *Scenario 1* is coming down when we consider the energy consumption, concluding that in the future the noosphere will not be sustainable. Also in this essay we discussed other minor criticisms (*i*,..., *iv*) to the hypothesis of singularity, taken from studies conducted by other thinkers. Our position does not deny the possibility of reaching $\Omega$ point, quite the contrary we propose that this would be possible if we are able to redesign the hardware that sup-ports the noosphere. A systemic model of the noosphere as simulated in *Scenario 2*, shows the possibility of a balance between the size of the noosphere and the amount of electrical energy available. Moreover, in *Scenario 3* we demonstrate how this equilibrium is even possible under the moderate effects of the energy paradigm. Finally, and doing a brainstorming session about how could be the hardware that implements the noosphere, we arrived to a very speculative computer architecture which we have called *N-computer*.

Despite all the criticism and arguments made, the main question remains open when our civilization will reach the $\Omega$ point or singularity?

## References

L.M.A. Bettencourt, J. Lobo, D. Helbing, C. Kuhnert, G.B. West. 2007. Growth, innovation, scaling, and the pace of life in cities. PNAS 104(17): 7301-7306.

F.S. Collins. 2006. The language of God. Free Press.

G. M. Church, Y. Gao, S. Kosuri. 2012. Next-generation dogital information storage in DNA. Science 337: 1628.

G. Egan. 1994. Permutation city. Millenium Orion Publishing Group.

J. F. Gantz (Ed.). 2008. The Diverse and Exploding Digital Universe. An IDC White paper.

H. de Garis. 2011. Femtocomputing. H+ Magazine. http://hplusmagazine.com/2011/11/01/ femtocomputing/

T.J. Garrett. 2011. Are there basic physical constraints on future anthropogenic emissions of carbon dioxide? Climate Change 104(3-4): 437-455.

S. Gavrilets, D.G. Anderson, P. Turchin. 2010. Cycling in the complexity of early societies. Cliodynamics: the Journal of Theoretical and Mathematical History 1(1): 58-80.

N. Goldman, P. Bertone, S. Chen, C. Dessimoz, E. M. LeProust, B. Sipos, E. Birney. 2013. Towards practical, high-capacity, low-maintenance information storage in synthesized DNA. Nature 494: 77–80.

A.F. Griffin. 2011. Emergence of fusion/fission cycling and self-organized criticality from a simulation model of early complex polities. Journal of Archaeological Science 38: 873-883.

B. Goertzel. 2011. From DNA Computing to Femtocomputing? H+ Magazine. http://hplusmagazine.com/2011/11/23/from-dna-computing-to-femtocomputing/

D.H. Good, R. Reuveny. 2009. On the collapse of historical civilizations. Amer. J. Agr. Econ. 91(4): 863-879.

L. Hagerty. 2000. The Spirit of the Internet: Speculations on the Evolution of Global Consciousness. Matrix Masters: 1-240.

S. Hameroff. 1998. Quantum computation in brain microtubules? The Penrose-Hameroff 'Orch OR' model of consciousness. Phil. Trans. R. Soc. Lond. A. 356: 1869-1896.

M. Heim. 1993. The Metaphysics of Virtual Reality. Oxford University Press.

J. Huebner. 2005. A possible declining trend for worldwide innovation. Technological Forecasting & Social Change, October: 980-986.

G. Karakatsanis. 2010. Modelling the energy dynamics of human civilizations within the theory of cultural evolutionism. Advances

in Energy Studies 2010, 7[th] Biennial International Workshop, Barcelona, Spain, pp. 1-10.

R. Kurzweil. 2005. The singularity is near. Viking: 1-652.

R. Lahoz-Beltra. 2004. Bioinformatica. Simulacion, vida artificial e inteligencia artificial. Ediciones Diaz de Santos, Madrid. (Trans.: in Spanish).

R. Lahoz-Beltra. 2008. ¿Juega Darwin a los dados? Nivola (Trans.: in Spanish).

E. Le Roy. 1927. L'exigence idéaliste et le fait de l'évolution. Boivin & Cie.

G.S. Levit. 2000. The Biosphere and the Noosphere Theories of V. I. Vernadsky and P. Teilhard de Chardin: A Methodological Essay. International Archives on the History of Science/Archives Internationales D'Histoire des Sciences 50 (144): 160-176.

A.J. Lotka. 1925. Elements of physical biology. Williams and Wilkins.

J.E. Lovelock. 1979. Gaia A new look at life on Earth. Oxford University Press.

T. Modis. 2003. The limits of complexity & change. THE FUTURIST (May-June): 26-32.

T. Modis. The singularity myth. Technological Forecasting & Social Change 73(2).

J.C. Nuño, J. de Vicente, J. Olarrea, P. Lopez, R. Lahoz-Beltra. 2010. Evolutionary Daisyworld models: A new approach to studying complex adaptive systems Ecological Informatics 5(4): 231-240.

C. Recio Rincon, P. Cordero, J. Castellanos, R. Lahoz-Beltra. 2013. A new method for the binary encoding and hardware implementation of metabolic pahtways. International Journal Information Theories and Applications, in press.

D. Ronfeldt, J. Arquilla. 2008. From Cyberspace to the Noosphere: Emergence of the Global Mind. New Perspectives Quarterly 17(1):18-25.

R.G. Snapper. Language of the Noosphere. http://www.nyu.edu/classes/ keefer/com/snap1.html

D. Shenk. 1997. Data Smog. New York: Harper Collins.

P. Teilhard de Chardin. 1959. The Phenomenon of Man. New York: Harper and Row.

P. Teilhard de Chardin. 1964. The Future of Man (1964) Image.

S. Ulam. 1958. Tribute to John von Neumann. Bulletin of the American Mathematical Society 64: 5.

V. I. Vernadsky. 2004. The Biosphere and the Noosphere. Airis Press, Moscow [Transl.: in Russian].

V. Vinge. 2013. Vernor Vinge on the Singularity. San Diego State University. http://mindstalk.net/vinge/vinge-sing.html

V. Volterra. 1926. Variazioni e fluttuazioni del numero d'individui in specie animali conviventi. Mem. R. Accad. Naz. dei Lincei. Ser. VI. 2.

J. Watson, J.E. Lovelock, J.E. 1983. Biological homeostasis of the global environment: the parable of Daisyworld. Tellus 35B (4): 286–9.

S.L. Weber. 2001. On homeostasis in Daisyworld. Climatic Change 48: 465-485.

L. White. 1959. The evolution of culture: the development of civilization to the fall of Rome. McGraw-Hill.

S. Wolfram. 2002. A new kind of science. Wolfram Media.

## Appendix

The numerical model simulations were performed on Scilab 5.4.1 environment.

### *Scenario 1*

```
//Noosphere equations
//Exponential model

function [w] = f(t,y)
w(1) = y(1)*k1;
endfunction

k1 =2;
t0 = 0; y0 = [1000];
t = [0:0.01:17];
y = ode(y0,t0,t,f);
xset('window',1)
y1 = y(1,:);
clf;
plot2d(t+2013,y1,style=1);
xtitle('Noosphere collapse','t','EXABYTES');

//end script
```

*Scenario 2*

```
 //Lotka-Volterra equations
 //y1 = prey-energy population, y2 = preda-
tor-noosphere population
 //dy1/dt = y1*(k1-k3*y2), dy2/dt = y2*(-
k2+k3*y1)
 //Use k1 = 1, k2 = 10, k3 = 1 for 0 < t <
100
 //y1(0) = 30, y2(0) = 80

 function [w] = f(t,y)
 w(1) = y(1)*( k1-k3*y(2));
 w(2) = y(2)*(-k2+k3*y(1));
 endfunction

 k1 = 1; k2 = 10; k3 = 1;
 t0 = 0; y0 = [0.02;1];
 t = [0:0.01:100];
 y = ode(y0,t0,t,f);
 y1 = y(1,:); y2 = y(2,:);
 mtlb_subplot(2,2,1);plot2d(t,y1);xtitle('Ene
rgy','t','y1');
 mtlb_subplot(2,2,2);plot2d(t,y2);xtitle('Noo
sphere  population','t','y2');
 mtlb_subplot(2,2,3);
 plot2d(y1,y2);xtitle('phase    portrait','y1
ENERGY','y2 NOOSPHERE');

 //end script
```

*Scenario 3*

```
//Lotka-Volterra equations under Energy Pa-
radigm dynamics
//y1 = prey-energy population, y2 = preda-
tor-noos population
//dy1/dt = y1*(b-c*y2), dy2/dt = y2*(-
d+c*y1)
//Use k1 = 1, k2 = 10, k3 = 1 for 0 < t <
100
//y1(0) = 30, y2(0) = 80

function [w] = f(t,y)
w(1) = y(1)*( k1-k3*y(2))- 0.01 * y(1)^2;
w(2) = y(2)*(-k2+k3*y(1));
endfunction

k1 = 1; k2 = 10; k3 = 1;
t0 = 0; y0 = [0.02;1];
t = [0:0.01:100];
y = ode(y0,t0,t,f);
y1 = y(1,:); y2 = y(2,:);
clf
mtlb_subplot(2,2,1);plot2d(t,y1);xtitle('Ene
rgy','t','y1 ENERGY');
mtlb_subplot(2,2,2);plot2d(t,y2);xtitle('Noo
sphere','t','y2 NOOSPHERE');
mtlb_subplot(2,2,3);
plot2d(y1,y2);xtitle('phase    portrait','y1
ENERGY','y2 NOOSPHERE');

//end script
```